\documentclass[10pt]{article}
\usepackage[explicit]{titlesec}
\usepackage{hyphenat}
\usepackage{ragged2e}
\RaggedRight

\usepackage[T1]{fontenc}
\usepackage{amsmath, amsthm}
\usepackage{graphicx}
\usepackage{url}
\usepackage{hyperref}
\usepackage{soul}
\setul{.6pt}{.4pt}
\usepackage{geometry}
\usepackage{fancyhdr}
\usepackage[labelfont={footnotesize,bf} , textfont=footnotesize]{caption}
\makeatletter
\renewcommand\tagform@[1]{\maketag@@@ {\ignorespaces {\footnotesize{\textbf{Equation}}} #1.\unskip \@@italiccorr }}
\makeatother
\titleformat{\section}
  {\normalfont}{\thesection}{1em}{\MakeUppercase{\textbf{#1}}}
\titlespacing\section{0pt}{0pt}{-10pt}
\titleformat{\subsection}
  {\normalfont}{\thesubsection}{1em}{\textit{#1}}
\titlespacing\subsection{0pt}{0pt}{-8pt}

\makeatletter
\newcommand\sixteen{\@setfontsize\sixteen{17pt}{6}}
\renewcommand{\maketitle}{\bgroup\setlength{\parindent}{0pt}
\begin{flushleft}
\sixteen\bfseries \@title
\medskip
\end{flushleft}
\textit{\@author}
\egroup}
\makeatother

\usepackage[sort&compress]{natbib}
\makeatletter
\renewcommand\@biblabel[1]{\textbf{#1.}\hfill}
\makeatother

\bibpunct{}{}{,~}{s}{,}{,}
\def\mbf#1{\mathbf{#1}}

\title{Modeling and Simulation of Traffic on I-485 via Linear Systems and Iterative Methods}

\author{Dominic Kealoha, Fabiola Rojas, Xingjie Li \\ \medskip
Department of Mathematics and Statistics, University of North Carolina at Charlotte, Charlotte, NC \\ \medskip
Students: dkealoha@charlotte.edu, frojas4@charlotte.edu  \\
Mentor: xli47@charlotte.edu
}

\begin{document}

\vspace*{.01 in}
\maketitle
\vspace{.12 in}

\section*{abstract}
Iterative methods such as Jacobi, Gauss-Seidel, and Successive Over-Relaxation (SOR) are fundamental tools in solving large systems of linear equations across various scientific fields, particularly in the field of data science which has become increasingly relevant in the past decade. Iterative methods' use of matrix multiplication rather than matrix inverses makes them ideal for solving large systems quickly. Our research explores the factors of each method that define their respective strengths, limitations, and convergence behaviors to understand how these methods address drawbacks encountered when performing matrix operations by hand, as well as how they can be used in real world applications. After implementing each method by hand to understand how the algorithms work, we developed a Python program that assesses a user-given matrix based on each method's specific convergence criteria. The program compares the spectral radii of all three methods and chooses to execute whichever will yield the fastest convergence rate. Our research revealed the importance of mathematical modeling and understanding specific properties of the coefficient matrix. We observed that Gauss-Seidel is usually the most efficient method because it is faster than Jacobi and doesn't have as strict requirements as SOR, however SOR is ideal in terms of computation speed. We applied the knowledge we gained to create a traffic flow model of the I-485 highway in Charlotte. After creating a program that generates the matrix for this model, we were able to iteratively approximate the flow of cars through neighboring exits using data from the N.C. Department of Transportation. This information identifies which areas are the most congested and can be used to inform future infrastructure development.

\section*{keywords}
Iterative Methods; Convergence; Spectral Radius; Traffic Flow Network

\vspace{.12 in}






\section{Introduction}\label{sec:intro}

There are various ways to solve a system of linear equations, notably direct methods and iterative methods. During our 10 weeks of undergraduate research, we closely examined three iterative methods intended for solving systems of linear equations, those being the Jacobi Method, Gauss-Seidel Method, and Successive Over-Relaxation Method. From our time studying these models we were able to determine their differences in general efficiency and when they might be most effectively applied. From this point we developed a program based in Python in which a user can input a given square matrix and desired accuracy and find its solution using the iterative method that will return quickest results.   We developed a generalized process using flow networks for the translation of real world data into a matrix form that guarantees the effective use of the three iterative methods we studied. Using the average annual daily traffic (AADT) data values on each exit of the I-485 highway in the city of Charlotte, NC, we were able to successfully approximate the amount of traffic on every segment of road. We have opted to use iterative methods for this modeling because of their utilization of matrix multiplication which allows significantly higher computational speed making them ideal for exceptionally large data sets.

Applying iterative methods to determine the AADT of I-485 has proven to be insightful. Our data highlights neglected infrastructure in regards to traffic flow in exits near the UNCC campus, indicating a rapidly increasing student population. This claim is supported by reduced AADT values in June and July where the majority of students return home for Summer break. The Charlotte Department of Transportation (CDOT) and North Carolina Department of Transportation (NCDOT), who in collaboration are responsible for all construction projects on I-485, have shown a focus on road construction in the southern regions of I-485. Our findings show that exits near the UNCC campus on I-485 had the highest AADT values,  which we did not expect given the high density of Uptown as well as it being the nation's second largest banking center. These findings reflect the increasing population of UNCC as well as a need for more efficient infrastructure in the general area. The modeling of the average traffic along I-485 can play a significant role in determining what regions of I-485 are receiving less or more support in terms of infrastructure, from the CDOT and NCDOT. The model we created is also flexible, and in the future we hope to address more issues in Charlotte infrastructure such as the effectiveness of Duke Energy's current power supply grid in Charlotte.

In the following we will provide an overview of our research and provide our findings. First, we provide an introduction and some important prerequisite information involving our general methodology. We also review the three iterative methods that we studied and observe their unique analytical properties. Next, we elaborate on our Python program and how we implemented the algorithms to promote efficiency, and discuss applying what we've learned to approximate traffic values on I-485 using iterative methods and discussing the AADT approximations. We also explain our method for making any similar matrix fit the convergence criteria for the three iterative methods. Finally, we have a general conclusion discussing our findings, the importance of their application, and possible applications that we would like to consider in the future.

\section{Method}\label{sec:method}
In this section, we will review three iterative methods for solving linear systems according to the tutorials \cite{web1,web3,web4}, as well as the details on converting a traffic network to a linear system by following the textbook \cite{book}.

\subsection{Linear Systems of Equations and Direct Methods}\label{subsec:direct_methods}
Iterative methods have proven to be particularly effective in solving large systems of equations with impressive memory efficiency and solution accuracy. To help us understand why this is, the following is a brief general overview of concepts involving linear systems of equations, and defining iterative methods as a whole while acknowledging their strengths and weaknesses.

Linear systems share the following general format where we define $A$ as the coefficient matrix, $\mbf{x}$ as the vector of unknown variables that we intend to solve for, and $\mbf{b}$ as our vector of constants.
\begin{equation}\label{gen_LinearSys}
A\mbf{x}  = \mbf{b}.
\end{equation}
This system of equations is able to be translated into a matrix representation. The general steps for this process involve pulling all of the equation coefficients into matrix $A$ and a right hand side vector $\mbf{b}$, and the unknown solution $\mbf{x}$ could be solved by sequentially  eliminating unknown variables. 
For instance, the following system of equations:
\begin{align*}
5x_1 - 2x_2 + 3x_3 &= -1, \\
-3x_1 + 9x_2 + 1x_3 &= 2, \\
-2x_1 - 1x_2  -7x_3 &= 3
\end{align*}
share a matrix representation of:
\begin{equation*}
\begin{array}{ccc}
\begin{bmatrix}
5 & -2 & 3 \\
-3 & 9 & 1 \\
-2 & -1 & -7
\end{bmatrix}
\times
\begin{bmatrix}
x_1 \\
x_2 \\
x_3
\end{bmatrix}
=
\begin{bmatrix}
-1 \\
2 \\
3
\end{bmatrix}.
\end{array}
\end{equation*}
By Gaussian elimination, we convert the above system into
\[
\begin{bmatrix}
5 &-2 & 3\\
0 & 39 & 14\\
0 & 0 &\frac{-335}{13}
\end{bmatrix}
\times
\begin{bmatrix}
x_1 \\
x_2 \\
x_3
\end{bmatrix}
=
\begin{bmatrix}
-1 \\
7 \\
\frac{190}{13}
\end{bmatrix}
\]
and hence we can get the solution.

To perform the direct method, the matrix $A$ in \eqref{gen_LinearSys} needs to be invertible.



\subsection{Iterative Methods}\label{subsec:ite_methods}
Other than the direct methods, we can also solve $A\mbf{x}=\mbf{b}$ by iterative methods.
 In this subsection, we review three common iterative methods for solving the linear system in \eqref{gen_LinearSys} The core of our research revolved around the study of iterative methods for their potential use in solving linear systems arising from various real world applications.

Let $A_{ij}$ denote an entry in the matrix $A$ at row $i$ and column $j$ and let $k$ denote the current iteration number. 
We are able to approximate a solution by introducing an initial guess for the solution $\mbf{x}$ that we then repeatedly refine through multiple iterations using one of the iterative algorithms. For the sake of consistency, during our study we opted to always set our initial guess $\mbf{x}^{(0)}$ to be a zero vector for more definitive results in comparing the performances of various algorithms. At the $k$-th iteration, the current iterative value is denoted by $\mathbf{x}^{(k)}$. The program we created utilizes the norm of the residual to track how close the iterations are getting to the true solution. Recall that for a linear system $A\mbf{x}=\mbf{b}$, the residual for the $k$-th iterative solution  is defined as $\mbf{r}^{(k)} := \mbf{b}-A\mbf{x}^{(k)}$. We know we are approaching the convergence point once our estimated quantity satisfies the pre-determined error threshold $\eta$, that is $\|\mbf{r}^{(k)}\|_{2}<\eta$.  After each iteration the difference between $\mbf{b}$ and $A\mbf{x}^{(k)}$ gets smaller and smaller, meaning the $k$-th approximation is getting closer and closer to the exact solution $\mbf{x}$.

\paragraph{The Jacobi method.} 	First, we studied the Jacobi iterative method. The algorithm is as follows:
\begin{equation}\label{Jacobi}
\mbf{x}_i^{(k)} = \frac{1}{A_{ii}} \left( \sum_{j=1, j \neq i}^n (-A_{ij} \mbf{x}_j^{(k-1)}) + \mbf{b}_i \right), \quad i = 1, 2, \ldots , n.
\end{equation}

\paragraph{The Gauss-Seidel method.} Next, we studied the Gauss-Seidel algorithm:
\begin{equation}\label{GS_method}
\mbf{x}_i^{(k)} = \frac{1}{A_{ii}} \left(  -\sum_{j=1}^{i-1} A_{ij} \mbf{x}_j^{(k)} - \sum_{j=i+1}^n (A_{ij} \mbf{x}_j^{(k-1)}) + \mbf{b}_i \right), \quad i = 1, 2, \ldots, n .
\end{equation}

\paragraph{The successive overrelaxation method (SOR).}
The SOR method is similar to the Gauss-Seidel method (GS) and the update of $\mathbf{x}^{(k)}$  can be viewed as the combination of it previous value and GS value through a new weight parameter $\omega$:
\begin{equation}\label{SOR}
\mbf{x}_i^{(k)} = (1 - \omega) \mbf{x}_i^{(k-1)} + \frac{\omega}{A_{ii}} \left( \mbf{b}_i - \sum_{j=1}^{i-1} A_{ij} \mbf{x}_j^{(k)} - \sum_{j=i+1}^n A_{ij} \mbf{x}_j^{(k-1)} \right), \quad i = 1, 2, \ldots, n.
\end{equation}
Suppose that $A=D-L-U$, where $D$ is a diagonal matrix that contains the diagonal part of $A$, $-L$ and $-U$ are the lower and upper triangular parts of $A$, respectively.
Each algorithm then can be translated into a simpler matrix form:
\[
\mbf{x}^{(k)}= T \mbf{x}^{(k-1)}+\mbf{c},
\]
where $T$ is the iterative matrix and $\mathbf{c}$ is a constant vector:
\begin{itemize}
\item The Jacobi method:
\[
T_{\text{j}}:=D^{-1}\left(L+U\right),\quad
\mbf{c}_{\text{j}}:=D^{-1}\mbf{b}.
\]
\item The Gauss-Seidel method:
\[
T_{\text{g}}:=\left(D-L\right)^{-1} U,\quad \mbf{c}_{\text{g}}:=\left(D-L\right)^{-1} \mbf{b}.
\]
\item The SOR method:
\[
T_{\omega}:=\left(D-\omega L\right)^{-1}\left((1-\omega)D+\omega U\right),\quad
\mbf{c}_{\omega}:= \omega \left(D-\omega L\right)^{-1}\mbf{b}.
\]
\end{itemize}

\subsection{Convergence criteria}\label{subsec:converg}
We began our research by studying the algorithms and convergence criteria for each iterative method. Each method shares the same sufficient and necessary criteria about the working matrix in order for the method to converge to a solution, that is the matrix $T$ must have a spectral radius less than 1:
\[
\rho(T)<1.
\]
Here, $\rho(T)$  is defined as the radius of the smallest circle in the complex plane that contains all eigenvalues of matrix $T$.
We know the matrix will converge if and only if it meets these conditions. For the SOR method, another necessary condition for convergence is that $0<\omega<2$. 
Each method also has some more conditions regarding the structure of the matrix that are sufficient to determine convergence. For Jacobi, if $A$ is diagonally dominant, we know that $T_{\text{j}}$ will converge. For Gauss-Seidel, the method is convergent if $A$ is symmetric and positive-definite. For SOR, $T_{\omega}$ is convergent if $A$ is symmetric, positive-definite, and tridiagonal. For each method, the smaller $\rho(T)$ is, the faster the convergence speed is.

The program that we developed first checks the norm of residual after the estimated number of iterations are completed, and if the current residual norm is still above the threshold, i.e., $\|\mathbf{r}^{(k)}\|\ge \eta$, then the iterations will be continued. When the residual falls under the threshold, the program displays the final solution. It is important to note that this is an approximation, and we can never get the exact solution using iterative methods. In theory, iterations can continue infinitely. The purpose of introducing an error threshold is to stop iterations at a point where the residuals are sufficiently small.

To save the computational cost and avoid checking the residual norm every iteration, we can approximate the total number of iterations according to a given threshold $\eta$. For each method, we have the following error estimate of the iterative solution
\[
\|\mbf{x}-\mbf{x}^{(k)}\|\le \frac{\|T\|^{k}}{1-\|T\|}\|\mbf{x}^{(1)}-\mbf{x}^{(0)}\|,
\]
therefore, for residual $\mbf{r}^{(k)}$ we have
\[
\begin{split}
\|\mbf{r}^{(k)}\|=\|\mbf{b}-A\mbf{x}^{(k)}\|
=&\|A\left(\mbf{x}-\mbf{x}^{(k)}\right)\|\\
\le& \|A\|\times  \|\mbf{x}-\mbf{x}^{(k)}\|\le \|A\|\frac{\|T\|^{k}}{1-\|T\|}\|\mbf{x}^{(1)}-\mbf{x}^{(0)}\|.
\end{split}
\]
Consequently, for a given threshold $\eta$, the pre-determined total number of iterations is given by
\begin{equation}\label{pre-det-k}
\widehat{k}\ge \frac{\log\left(\frac{\eta (1-\|T\|) }{\|A\|\times \|\mbf{x}^{(1)}-\mbf{x}^{(0)}\|}\right)}{\log\left(\|T\|\right)}
= \frac{\log\left(\frac{\eta (1-\rho(T)) }{\|A\|\times \|\mbf{x}^{(1)}-\mbf{x}^{(0)}\|}\right)}{\log\left(\rho(T)\right)},
\end{equation}
where $\rho(T)$ is the spectral radius of the matrix $T$.



For SOR method, the convergence speed depends on the optimal weight: $\omega^*=\mathrm{argmin}_{0<\omega<2}\, \rho(T_{\omega})$, which is usually difficult to compute. However, if $A$ is a positive definite and tridiagonal matrix, then the optimal $\omega^*$ for SOR is
\begin{equation}\label{opt_SOR}
\omega^* = \frac{2}{1 + \sqrt{1-\rho\left(T_\text{j}\right)^2}},
\end{equation}
where $\rho\left(T_\text{j}\right)$ is the spectral radius of the corresponding Jacobi method, and the optimal spectral radius for SOR is $\rho\left(T_{\omega}\right)=\omega^*-1$.

\subsection{Comparing Methods}
Typically, we solve these matrices using direct methods via Gaussian elimination and row reduction. 
There are two main issues from using direct methods. Firstly, they are quite inefficient in general, and require an abundance of  unnecessary computational memory resources and require far more time to reach a solution. When it comes to larger matrices containing data that can not be represented by just a few variables, or a matrix that is too large and sparse to find a solution in a reasonable amount of time, this issue becomes especially apparent. Secondly, there are cases in which direct methods are unable to reach a solution entirely, like in cases where a matrix might be non-invertible. These are two issues that iterative methods do not share.

Iterative methods really shine when it comes to real-world application as they can be applied very easily to large or sparse matrices. These methods are very time efficient, reaching solutions far faster than direct methods. They are far less taxing on memory storage and computational resources. This is since we only need specific information from large matrices or sparse matrices. These methods also have the ability to solve non-invertible matrices.

One of the key differences that allow for iterative methods to be computation- and memory-efficient is the use of matrix multiplication, which negates the reliance that direct methods have on finding an inverse matrix to find a solution.

\subsection{Flow Networks and Traffic Modeling}\label{subsec:FlowNetwork}

\begin{figure}[ht!]
\centering
\includegraphics[width=0.4\textwidth]{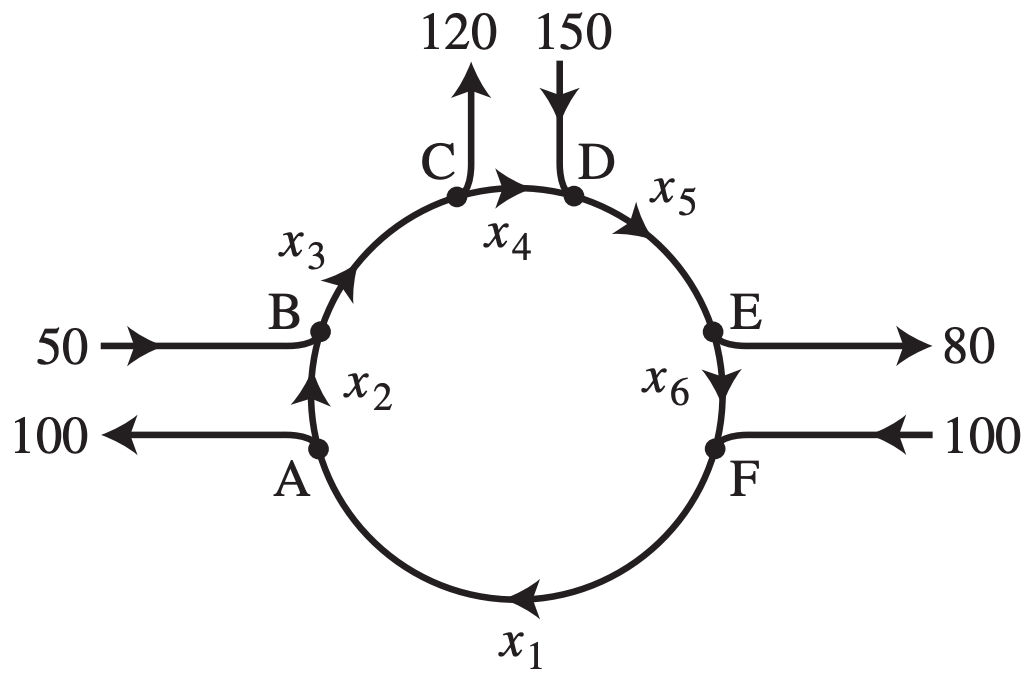}
\caption{A graphical illustration of a simple traffic flow network. }\label{Fig:FlowNet}
\end{figure}

In our search for real world applications of iterative methods, we came across flow networks. Flow network diagrams are a great way to visualize the data in linear systems. These diagrams are composed of two primary components. The first component being junctions which are labeled with capital letters in Figure~\ref{Fig:FlowNet} .

Note that we can define a junction as any point or \textit{node} with values flowing in or out. The second main component are \textit{branches}, which are described by the directional flow (indicated by directional arrows) and flow amount (indicated by the number values, or placeholder variables). We can define branches as the connections between nodes.

We are able to use each junction's input and output values (defined by the branches entering and exiting each node) to create a linear equation. The input values belong to the left hand side of the equation, and the output is represented by the right. This can be done for every node in a diagram. For the illustration network in Figure~\ref{Fig:FlowNet}, the associated linear system reads:
\[
\begin{aligned}
& A: \quad x_1 = x_2+100, \\
& B: \quad x_2+50 = x_3,\\
& C: \quad x_3= 120+x_4,\\
& D: \quad x_4+150 = x_5,\\
& E: \quad X_5=80+x_6,\\
& F: \quad x_6 +100 = x_1.
\end{aligned}
\]

\section{Numerical Simulation}
\subsection{Modeling Highway I-485}
\begin{figure}[ht!]
\centering
\includegraphics[width=0.4\textwidth]{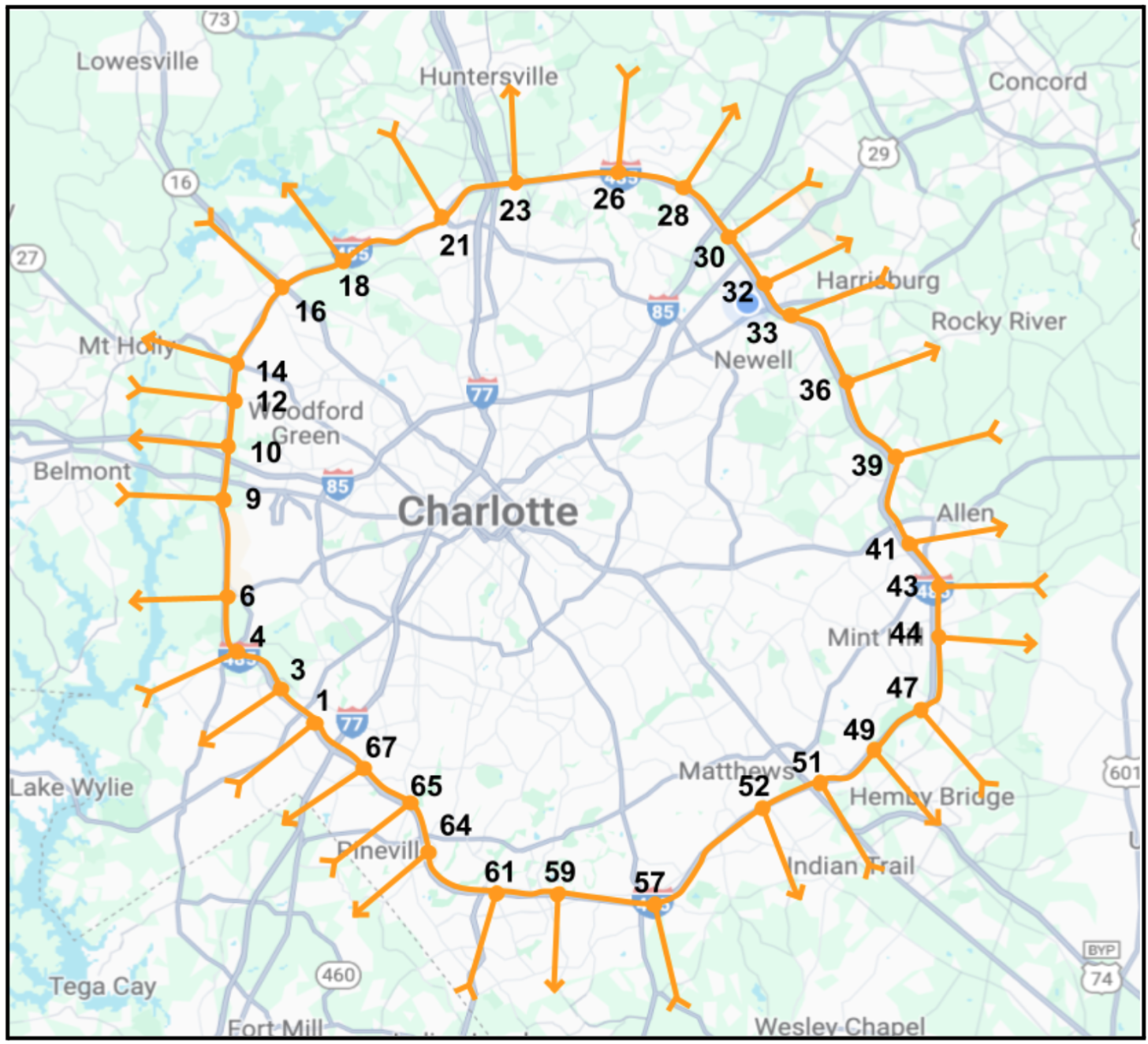}
\caption{A traffic Network-Flow of highway I-485 in the city of Charlotte.}
\label{Fig:I485}
\end{figure}

The circular flow network discussed in Subsection~\ref{subsec:FlowNetwork} inspired us to think about modeling highway I-485 in Charlotte, as demonstrated in Figure~\ref{Fig:I485}.

\vspace{-0.25 cm}

\vspace{0.2 cm}
Visually, it is reminiscent of a circular highway with multiple exits along the circumference. We also observed a notable pattern that encouraged us to explore using data from I-485. This example diagram alternates input and output data at each branch, resulting in the $1$, $-1$, $0$ pattern in the matrix, which can be recast as 32-by-32 matrix
\begin{equation}\label{I485_matrix}
A=\begin{bmatrix}
    1 & -1 & 0 & \dots & 0 &0 &0\\
0 & 1 & -1& \dots & 0 &0 &0\\
0 & 0 & 1 &\dots &0 &0 &0\\
\vdots &\vdots &\vdots &\ddots&\vdots&\vdots&\vdots\\
0 & 0 & 0 &\dots  &1 &-1 &0\\
0 & 0 & 0 &\dots &0 &1 &-1\\
-1 & 0 & 0 &\dots &0 &0 &1
\end{bmatrix}_{32\times 32}\quad
\end{equation}

\textit{However}, the $32\times 32$ matrix $A$ is a circular matrix, hence $A$ has a simple eigenvalue of zero. Therefore, any iterative matrix $T$ resulting from this flow network has a spectral radius of 1, which makes it ineligible to solve with one of the three iterative methods. To solve this problem, we utilized some properties regarding the eigenvalues and eigenvectors of the matrix. Meanwhile, because that $A$ had a simple eigenvalue $\lambda=0$ and a corresponding eigenvector of $e=\begin{bmatrix}1 & 1\dots 1&1\end{bmatrix}^{\mathrm{T}}$, by definition, $A\times e =\lambda\times e$, so $A \begin{bmatrix}1 & 1\dots 1&1\end{bmatrix}^{\mathrm{T}} = \mathbf{0}$. From here we remove the last column of $A$ in order to remove the simple zero eigenvalue, resulting in a new 32x31 matrix $\tilde{A}$. To modify the linear system accordingly, we must then also remove the last row of solution $\mbf{x}$  as it corresponds to that last column. Now we have defined a new linear system $\tilde {A}\tilde{\mbf{x}}=\mbf{b}$, which is not a square linear system. The next step is to make the equation into a normal equation
\[
\tilde{A}^{\mathrm{T}} \tilde {A}\tilde{\mbf{x}}=\tilde{A}^{\mathrm{T}}\mbf{b},
\]
resulting in a $31\times 31$ matrix $\tilde{A}^{\mathrm{T}} \tilde {A}$ and $31\times 1$ vectors $\tilde{A}^{\mathrm{T}} \mbf{b}$ as well as the solution $\tilde{\mbf{x}}$.

Now, we have a square linear system, we can enter it into our code \cite{github} to calculate the values for $\tilde{\mbf{x}}$. The calculations were made without the 32nd and final entry of the original $\mbf{x}$ vector, denoted by $c$ so we must approximate a value for $c$ and adjust $\tilde{\mbf{x}}$ to compensate for the missing element. We decided that the \textit{median} of the $\tilde{\mbf{x}}$ values was the most reasonable approximation for $c$. The final step is to append the 32nd entry $c$ back into $\tilde{\mbf{x}}$, and then increment each element of  $\tilde{\mbf{x}}$ by the value of $c$. For the right hand side vector $\mbf{b}$, we use the real world data from the North Carolina Department of Transportation website \cite{web2}. Consequently, the result of the linear system is the final approximation of the annual average daily traffic volume on each exit of I-485.

\subsection{Results}
The goal of our application of iterative methods to real world data was to accurately and rapidly approximate the values of the unknown $\mbf{x}$ vector for average traffic volume on each exit of I-485. After the manipulation of the traffic matrix to meet convergence criteria, we were able to upload the data to our program to be solved using iterative methods. SOR returned the smallest spectral radius, indicating that it would converge the fastest, providing us with the quickest results just as we were hoping for. We tested out each method with an error threshold of $\eta=0.001$ to see just how different the convergence rates were, and found that SOR was nearly 40 times faster than Jacobi, a staggering result. The approximations were computed remarkably fast, and since we were able to adjust the matrix to fit the criteria for SOR (having the fastest convergence rate of the three methods) our calculations are executed in a fraction of a second. The results are summarized in Table~\ref{table1}.

\begin{figure}[htp!]
\centering
\includegraphics[width=0.55\textwidth]{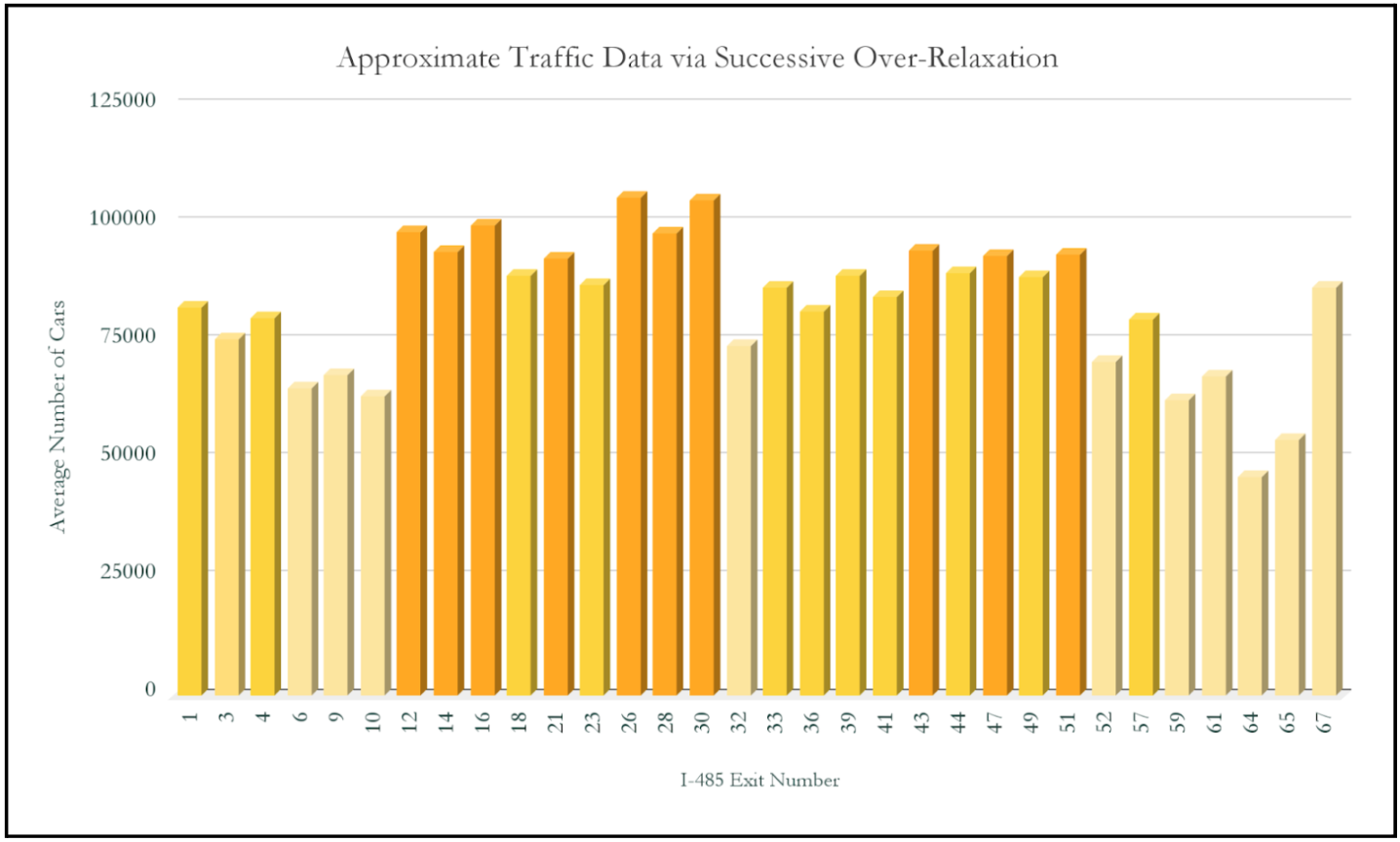}
\caption{The annual average traffic volume on each exit of I-485 in the city of Charlotte. }\label{Fig:I485-results}
\end{figure}

\begin{table}[htp!]
\begin{center}
\begin{tabular}{ |c| c| c| }
\hline
Iterative Method &Spectral Radius& \# of iterations \\
 \hline
Jacobi & 0.9952 & 3822 \\
\hline
Gauss-Seidel & 0.9904 & 1511 \\
\hline
SOR & 0.8215 & 99\\
\hline
\end{tabular}
 \caption{Performance summary of iterative methods solving  traffic volume of each exit on highway I-485. \label{table1}}
 \end{center}
\end{table}
\vspace{-0.5 cm}
We summarize the traffic volume of each exit of I-485 in Figure~\ref{Fig:I485-results}. Each bar in the graph represents the approximated average number of cars found to be between each exit (i.e. column 1 represents the average number of cars from exit 1 to exit 3 and so forth).
\subsection{Discussion}
The information in Figure~\ref{Fig:I485-results} provides powerful insight for potential Charlotte infrastructure. This data indicates areas of greater population density where lack of alternative routes are the cause of traffic congestion, like around exit 26 and exit 30. The framework for the data showcases flexibility, as if we were to hypothetically shut down an arbitrary number of exits for constructions or in the case of an accident, we would be able to approximate the traffic values of the segments with a new matrix that excludes those exits.

The actual numbers returned by the program were quite unexpected and opened up many questions about the current traffic situation in Charlotte. We found that the most dense areas in terms of car traffic are around exits 26 to 30, which are in the northernmost section of the highway and of the city. It is known that Charlotte is one of the nation's biggest banking hubs, and most of these large employers are located in Uptown Charlotte, right in the center of the city and the highway loop. As Charlotte residents, we also know that a majority of these employees either live in the outskirts of the center city or further south in the more affluent neighborhoods. Although most people who live in these areas likely would not need to take I-485 to commute Uptown, we did expect higher values around exits 52 through 65, which actually turned out to be some of the lowest values on the graph. We are still not clear what causes this, but we can speculate a few things:
(1) It could be that taking I-485 simply is not the most efficient route for most daily activities, including driving to work, school, grocery and other retail stores, and parks or other areas for leisure. (2) It is also possible that public transportation is more efficient in these areas, relieving some of the automobile congestion. (3) We also know that there has been more than five years of ongoing construction expanding this southern portion of the highway. It is possible that a few years ago our approximated AADT values would have been much higher due to closed lanes and detours. Now that the construction work is several years in, we are probably seeing the results of the highway expansion---that it seems to be effective in its goal to improve the flow of traffic in the area.

On the other hand, the northernmost area of the city is experiencing a lot more traffic on I-485---almost double. We did not expect this because this part of the city is not as densely populated as the southern region. The most dense areas of North Charlotte are UNC Charlotte and the area surrounding Concord Mills, a popular shopping center. These areas see the most traffic due to people's daily commute to school, work, or to go shopping. UNC Charlotte falls closer to exit 33, however, which is not as packed as some of the previous exits. Furthermore, most students live either on campus or in apartments that are very close to the school. \textit{This leads us to wonder if the level of traffic in these areas has not so much to do with the population but with the quality of infrastructure in this region.} As shown with our data and prior knowledge of South Charlotte, even though it holds a large percentage of the population, its AADT values are relatively low. We can speculate that residents of North Charlotte likely do not have as effective alternatives to the highway in order to reach the more popular destinations, like work or grocery stores.






\section{Conclusions}
In the large majority of cases, iterative methods have shown themselves to be the optimal approach to solving linear systems, especially when it comes to large or sparse real world data sets. We have seen that out of the three methods that we studied, the Jacobi Method is the most widely applicable with its less strict convergence criteria, although taking a considerably higher amount of iterations to meet the given error threshold. The Gauss-Seidel Method showed higher convergence rates than Jacobi, while having stricter convergence criteria. In all cases, SOR proved to have the highest convergence rates of the three by a substantial amount, while also having the strictest convergence parameters including the weight parameter $\omega$ which must be between zero and two. If able to meet the criteria for SOR, it is the most optimal iterative method of the three taking a fraction of the time and resources that the other methods need.

In general, iterative methods have a superior convergence rate and are overall more efficient in computation and memory costs than direct methods, showcasing superior practicality due to their use of matrix multiplication. These methods meet a demand that is ever-present in the new age of data collection and processing. We were able to compare the effectiveness of the three methods directly after creating a flow network diagram to translate traffic values into a matrix form for iterative method application, using the traffic AADT data on highway I-485 in the city of Charlotte, North Carolina. After inserting the data into our program, and manipulating the I-485 matrix to fit the criteria of SOR using properties of eigenvalues and eigenvectors, we were able to showcase SOR's efficiency by approximating the solution to the total 32 highway exits in less than 100 iterations, and within a fraction of a second.

The approximations we obtained highlight areas of high traffic congestion in the greater Charlotte area (northern and eastern hemispheres of I-485) opening up potential insights for infrastructural improvements across the city. With the scalability of the model we have developed utilizing flow networks and our Python program, we can also approximate traffic values if we were to close any amount of exits on I-485. This would prove useful in cases of road construction, or in the case of accidents obstructing the roadways. Scalability is a major advantage for many different potential applications and with the use of flow networks, translation of real data has become more efficient and we look forward to inserting larger matrices into our program, and increasing the scale of our research.

In the future, we would like to apply what we have learned to a much larger system to observe the flow of electricity for Duke Energy's power grid and  highlight areas that might be lacking in sufficient power. We would also like to optimize our Python program to further streamline the process of reading new matrices for real world application with large datasets in  future projects.

\section{Acknowledgements}
The authors thank Dr. Xingjie Li for her continuous support and mentorship all throughout and outside of the research process. The authors are incredibly grateful for the opportunities that she has provided and funding support from the NSF Award DMS-1847770.



\section*{about the student authors}

Dominic Kealoha is a senior in UNC Charlotte's College of Computing and Informatics concentrating in AI and is expected to graduate with a bachelor's degree in the Spring of 2025. Dominic has interests in applied mathematics in the field of data science and aspires to find potential implementations of mathematics to contribute to the field of artificial intelligence. Specifically, Dominic hopes to find ways to optimize the methods of the sorting  and collection of training data for AI development.

Fabiola Rojas is a student at UNC Charlotte's School of Data Science, expected to graduate in the Spring of 2025. Her inclination towards mathematics and statistics as well as interests in computer science and social matters drew her to pursue this study of her community. She hopes to continue to combine her love for mathematics and passion for social causes in her future career as a data scientist.

This research project marks the authors' first venture into academic research. With only undergraduate-level coursework in mathematics and computer science, they successfully conducted a detailed study on the traffic flow I-485---a topic that emerged due to its personal relevance---while incorporating advanced mathematical concepts.

During the course of their study, the authors have developed valuable skills that they will continue to refine and apply throughout their academic careers. They have gained significant experience in independently conducting research, formulating relevant questions, and articulating their findings in a professional manner. Both Dominic and Fabiola are committed to seeking further STEM-related opportunities and making a broader impact through their work. They are eager to continue exploring and making contributions to the STEM field.

\section*{press summary}

This study focuses on iterative methods such as Jacobi, Gauss-Seidel, and Successive Over-Relaxation (SOR), and how they were utilized to model traffic flow on the I-485 highway in Charlotte, NC. These methods, known for their efficiency in solving large systems of linear equations, were implemented in a Python program developed by the students to assess traffic data retrieved from the N.C. Department of Transportation and model the daily average number of cars on any given segment of the highway I-485, which is the one of most important highways in Charlotte. The study revealed that the Gauss-Seidel and SOR methods, particularly the latter, offered faster convergence rates, providing rapid approximations of traffic flow. The results highlighted a significant congestion near exits close to the UNC Charlotte campus, suggesting that city resources are not sufficiently focused on the busiest areas. This research not only advances the application of mathematical modeling in traffic management but also underscores the need for targeted infrastructure improvements in high-traffic regions.


\begin{thebibliography}{}
\bibitem{github} Our codes and values for all exits on I-485, as well as other information expanding approximations are available at:

\url{https://github.com/fabiola-rojas/iterative-methods-research2024} 


\bibitem{web1} Chapter 5 Iterative Methods for Solving Linear Systems. Courtesy of the University of Pennsylvania,

\url{https://www.cis.upenn.edu/~cis5150/cis515-12-sl5.pdf}
(accessed Mar 2024).

\bibitem{book} Lay, D. C., Lay, S. R., and McDonald, J. (2016) \textit{Linear Algebra and Its Applications}, Pearson, Boston.

\bibitem{web2} NCDOT. N.C. Department of Transportation Annual Average Daily Traffic (AADT) Mapping Application, \textit{ArcGIS web application},

\url{ https://ncdot.maps.arcgis.com/apps/webappviewer/index.html?id=964881960f0549de8c3583bf46ef5ed4} 
(accessed Mar 2024).

\bibitem{web3} Section 7.3: the Jacobi and Gauss-Seidel Iterative Methods. Courtesy of the University of Notre Dame:

\url{https://www3.nd.edu/~zxu2/acms40390F12/Lec-7.3.pdf} (accessed Mar 2024).

\bibitem{web4} Section 7.4: Relaxation Techniques for Solving Linear Systems. Courtesy of the University of Notre Dame:

\url {https://www3.nd.edu/~zxu2/acms40390F13/Lec-7.4-5.pdf} (accessed Mar 2024).

\end{thebibliography}
\end{document}